\documentclass[twocolumn,showpacs,preprintnumbers,amsmath,amssymb,elsart]{revtex4}
\usepackage{mathrsfs}
\usepackage{graphicx}% Include figure files
\usepackage{dcolumn}% Align table columns on decimal point
\usepackage{bm}% bold math
\usepackage[center]{subfigure}

\begin{document}

%\preprint{APS/123-QED}

\title{Coherent control of light field with electromagnetically induced transparency in a dark state Raman coherent tripod system}% Force line breaks with \\

\author {YongYao Li$^{1,2}$}
\author{HuaRong Zhang$^{1}$}
\author{YongZhu Chen$^{1,3}$}
\author{JianYing Zhou$^{1}$}
 \email{stszjy@mail.sysu.edu.cn}
\affiliation{$^{1}$State Key Laboratory of Optoelectronic Materials
and Technologies,\\Sun Yat-sen University, Guangzhou 510275, China\\
$^{2}$Department of Applied Physics,South China Agricultural
University, Guangzhou 510642, China\\
$^{3}$School of Electro-Mechanic Engineering,Guangdong Polytechnic
Normal University, Guangzhou 510665, China}

\date{\today}% It is always \today, today,

\begin{abstract}
The coherent superposition of two-atomic levels induced by coherent
population trapping is employed in a standard $\Lambda$ type scheme
to form a tripod-like system. A weak probe pulse scanning across the
system is shown to experience a crossover from absorption to
transparent and then to amplifcation. Consequently the group
velocity of the probe pulse can be controlled to propagate either as
a subluminal, a standard, a superluminal or even a negative speed.
It is shown that the propagation behavior of the light field is
entirely determined and controlled by the initial states of the
coherent superposition.
\end{abstract}

\pacs{Valid PACS appear here}% PACS, the Physics and Astronomy
\maketitle

\section{INTRODUCTION}
Coherent population trapping (CPT)\cite{Shore,White,Scully2} and
Electromagnetically induced transparency
(EIT)\cite{Harris1997,Fleischhauer2005} are two different quantum
coherence processes which have led to the observation of many new
physics effects in quantum optics and atomic physics. CPT is a
preparation of atoms in a coherent superposition of ground and
metastable state sublevels, which is called dark state, for the
reason that this state is immune to excitation by a two-component
laser radiation under the two-photon resonance condition. EIT can
also be classified as a dark state. It can modify the optical
properties of a medium and results in making a resonant opaque
medium transparent, which has been observed in a wide range of
atomic, molecular, and condensed matter
systems\cite{Harris1990,Ham,Sera}. In the past two decades,
potential applications of EIT to coherent control for group
velocities of the probe field in quantum information and quantum
communication\cite{Xiao,Hau,Scully1,Lukin1}, to left-handed material
for negative refraction\cite{Oketel,Thomen,Fleischhauer2007} and to
enhanced nonlinear optics for frequency
conversion\cite{Harris3,Harris4} have been reported. The distinct
feature of EIT and CPT is that EIT causes not only a modification of
optical property of the medium but also the optical fields
themselves.

In this paper, we analyse the optical property of a five-level
atomic configuration illustrated in FIG. 1. Two coherent fields with
complex Rabi-frequencies $\Omega_{1}$ and $\Omega_{2}$ create a dark
state superposition of states $|1\rangle$ and
$|2\rangle$\cite{Shore}. By introducing a coherent superprosition in
the system, more degrees of freedom to the system can be added, then
we will gain a control over more physical
variables\cite{Kurizki,Paspalakis1}. The coherent superprosition of
the dark state is given as below:
\begin{eqnarray}
&&|D\rangle=\cos\theta|1\rangle-\sin\theta|2\rangle
\end{eqnarray}
Here
\begin{eqnarray}
&&\cos\theta={\Omega_{1}\over\sqrt{\Omega^{2}_{1}+\Omega^{2}_{2}}}\nonumber\\
&&\sin\theta={\Omega_{2}\over\sqrt{\Omega^{2}_{1}+\Omega^{2}_{2}}}
\end{eqnarray}
%By scanning a weak probe field
By scaning a weak probe field through the system, the system
exhibits a rich optical properties under different initial
conditions because of the Raman coherence of the dark state. We
found that not only double EIT channels\cite{Iftiquar} and
left-handed material\cite{Fleischhauer2007} but also a lot of other
effects such as Raman amplification\cite{Roman}, refractive index
enhancement\cite{Scully4}, slow light\cite{Xiao,Hau} and
superluminal light\cite{WLJ} can be simply achieved by
adiabatically\cite{Oreg,Kis} changing the initial state.

The mixing angle $\theta$ in Eq. (2) is determined by the relative
intensity of $\Omega_{1}$ and $\Omega_{2}$. This dark state plays
the role of the original ground state. Here the system resembles a
three-level $\Lambda$-type system. The probe field then experiences
a transition from the dark state $|D\rangle$ to the upper state
$|4\rangle$. A strong resonant coherent field $\Omega_{C}$ is
employed to couple states $|3\rangle$ and $|4\rangle$.

\begin{figure}
\includegraphics[scale=0.4]{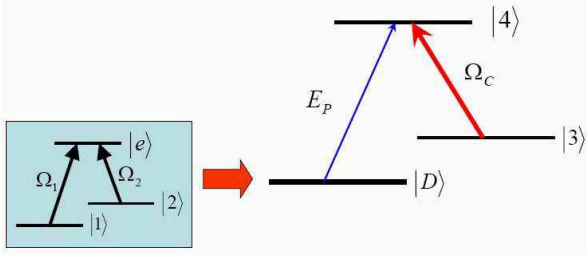}
\caption{\label{fig:wide}Energy-Level scheme of the system: two
coherent laser field with Rebi-frequency $\Omega_{1}$ and
$\Omega_{2}$ coupling of levels $|1\rangle$, $|2\rangle$ and
$|e\rangle$ create the dark state $|D\rangle$, which construct a
$\Lambda$ type scheme with $|3\rangle$ and $|4\rangle$. A probe
field  \emph{E} transits from dark state $|D\rangle$ to the upper
level $|4\rangle$ and a coupling field transits with $|3\rangle$ and
$|4\rangle$}
\end{figure}

The technique to establish the coherent preparation of a dark state
was realized in many systems\cite{Shore}. In our case, the atoms are
trapped in state $|1\rangle$ and state $|2\rangle$ with the
probability of $\cos^{2}\theta$ and $\sin^{2}\theta$ respectively.
The probability is decided by the mixing angle $\theta$, which can
be tuned adiabatically by changing the relative intensity of
$\Omega_{1}$ and $\Omega_{2}$. Because none of the atoms is prepared
at the state $|e\rangle$ and the probe field is weak, the
contribution from level $|e\rangle$ to the system can be neglected
safely. The energy levels ($|1\rangle$, $|2\rangle$, $|3\rangle$ and
$|4\rangle$) can be viewed as a tripod level
configuration\cite{Kis,Ham2,Mazets,Goren,David}. .

\section{EQUATIONS AND SOLUTIONS}
 The Hamiltonian of
this four-level system can be obtained in the rotating frame by
introducing the rotating-wave approximation:
\begin{eqnarray}
H=&&\sum_{i=1}^{4}\hbar\omega_{i}\sigma_{ii}-{\hbar\over
2}[\Omega_{1P}e^{-i\nu_{P}t}\sigma_{41}+\Omega_{2P}e^{-i\nu_{P}t}\sigma_{42}\nonumber\\
&&+\Omega_{C}e^{-i\nu_{C}t}\sigma_{43}+H.C]
\end{eqnarray}
Here $\omega_{i}$ is the eigenfrequency of the energy level and
$\sigma_{ij}=|i\rangle\langle j|$, Rabi-frequency
$\Omega_{1P}={\wp_{41}\mathscr{E}_{P}/ \hbar}$,
$\Omega_{2P}={\wp_{41}\mathscr{E}_{P}/ \hbar}$ and
$\Omega_{C}={\wp_{43}\mathscr{E}_{C}/ \hbar}$, where $\wp_{ij}$ is
the electric dipole moment from $|i\rangle$ to $|j\rangle$,
$\mathscr{E}_{P}$ and $\mathscr{E}_{C}$ is the slowly varying
amplitude of the probe field and the coupling field respectively.
Assume that the electric dipole moment
$\wp_{41}\thickapprox\wp_{42}=\wp$ and the value of $\wp$ is real,
which results that $\Omega_{1P}\approx\Omega_{2P}=\Omega_{P}$.

The evolution equation of density matrix reads:
\begin{eqnarray}
\dot{\rho}=-{i\over \hbar}[H,\rho]-[\Gamma,\rho]_{+}
\end{eqnarray}
The matrix element $\langle
i|\Gamma|j\rangle=\gamma_{i}\delta_{ij}$, where $\gamma_{i}$ is the
decay rate designating the population damping from the energy level
$|i\rangle$. To simplify, we assume that $\Omega_{1}$ and
$\Omega_{2}$ are real, and
$\gamma_{4}\gg\gamma_{3}\approx\gamma_{2}\approx\gamma_{1}\approx
0$.  With these approximations, the population of the atoms are
prepared at the energy levels $|1\rangle$, $|2\rangle$ and
$|3\rangle$ and the atoms are initially set to:
\begin{eqnarray}
&&\rho^{(0)}_{11}=\cos^{2}\theta,
\rho^{(0)}_{22}=\sin^{2}\theta\nonumber\\
&&\rho^{(0)}_{21}=\rho^{(0)}_{12}=-\cos\theta\sin\theta
\end{eqnarray}
The initial conditions indicate that $\cos^{2}\theta$ and
$\sin^{2}\theta$ denotes population of the atoms prepared at the
energy level $|1\rangle$ and $|2\rangle$ respectively. While
$-\cos\theta\sin\theta$ denotes the dark state Raman coherence
between levels $|1\rangle$ and $|2\rangle$.

The equations of density matrix elements in slowly varying amplitude
approximation are:
\begin{eqnarray}
&&\dot{\rho}_{41}=-(i\vartriangle_{1}+\gamma_{41})\rho_{41}+{i\over
2}\Omega_{P}\rho_{11}+{i\over 2}\Omega_{P}\rho_{21}+{i\over
2}\Omega_{C}\rho_{31}\nonumber\\
&&\dot{\rho}_{31}=-i(\vartriangle_{1}-\vartriangle_{C})\rho_{31}+{i\over
2}\Omega^{\ast}_{C}\rho_{41}\nonumber\\
&&\dot{\rho}_{42}=-(i\vartriangle_{2}+\gamma_{42})\rho_{42}+{i\over
2}\Omega_{P}\rho_{22}+{i\over 2}\Omega_{P}\rho_{12}+{i\over
2}\Omega_{C}\rho_{32}\nonumber\\
&&\dot{\rho}_{32}=-i(\vartriangle_{2}-\vartriangle_{C})\rho_{32}+{i\over
2}\Omega^{\ast}_{C}\rho_{42}
\end{eqnarray}

The off-diagonal decay rates $\gamma_{ij}$  are given by
$\gamma_{ij}={({\gamma_{i}+\gamma_{j})}/ 2}$ denoting the total
coherence relaxation rates between states $|i\rangle$ and
$|j\rangle$. The detuings are defined as
$\vartriangle_{1}=\omega_{41}-\nu_{P}$,$\vartriangle_{2}=\omega_{42}-\nu_{P}$
, $\vartriangle_{C}=\omega_{43}-\nu_{C}$. As shown in the system,
one finds that:
\begin{eqnarray}
\vartriangle_{1}-\vartriangle_{2}=\omega_{21}
\end{eqnarray}
The steady-state solutions of $\rho_{41}$ and $\rho_{42}$ are given
by:
\begin{eqnarray}
&&\rho_{41}={{{i\over
2}\Omega_{P}(\cos^{2}\theta-\cos\theta\sin\theta)(\vartriangle_{1}-\vartriangle_{C})}\over{(i\vartriangle_{1}+\gamma_{41})
(\vartriangle_{1}-\vartriangle_{C})-i\Omega^{2}_{C}/4}}\nonumber\\
&&\rho_{42}={{{i\over
2}\Omega_{P}(\sin^{2}\theta-\cos\theta\sin\theta)(\vartriangle_{2}-\vartriangle_{C})}\over{(i\vartriangle_{2}+\gamma_{42})
(\vartriangle_{2}-\vartriangle_{C})-i\Omega^{2}_{C}/4}}
\end{eqnarray}
The polarization $\mathscr{P}$ in the slowly varying frame and the
susceptibility $\chi$ are related to each other with the equation
\cite{Scully3}:
\begin{eqnarray}
\mathscr{P}=\epsilon_{0}\chi(\omega)
\mathscr{E}_{P}=2N(\wp_{14}\rho_{41}+\wp_{24}\rho_{42})
\end{eqnarray}
where \emph{N} is the total number of atoms.

The expression of the susceptibility is given as below:
\begin{eqnarray}
\chi(\omega,\theta)=&&iK[{{f(\theta)(\vartriangle_{1}-\vartriangle_{C})}\over{(i\vartriangle_{1}+\gamma_{41})
(\vartriangle_{1}-\vartriangle_{C})-i{\Omega^{2}_{C}/4}}}\nonumber\\
&&+{{g(\theta)(\vartriangle_{2}-\vartriangle_{C})}\over{(i\vartriangle_{2}+\gamma_{42})
(\vartriangle_{2}-\vartriangle_{C})-i{\Omega^{2}_{C}/4}}}
\end{eqnarray}
Where $K={N\wp^{2}/\epsilon_{0}\hbar}$, and
$f(\theta)=\cos^{2}\theta-\cos\theta\sin\theta$,
$g(\theta)=\sin^{2}\theta-\cos\theta\sin\theta$

Assume that $\gamma_{41}\approx\gamma_{42}=\gamma$ and
$\vartriangle_{C}=0$, the real part and imaginary part of
susceptibility are given as below:
\begin{eqnarray}
Re[\chi]=&&K[f(\theta){\vartriangle_{1}(\vartriangle^{2}_{1}-\Omega^{2}_{C}/4)\over{\gamma^{2}\vartriangle^{2}_{1}
+(\vartriangle^{2}_{1}-\Omega^{2}_{C}/4)^{2}}}\nonumber\\
&&+g(\theta){\vartriangle_{2}(\vartriangle^{2}_{2}-\Omega^{2}_{C}/4)\over{\gamma^{2}\vartriangle^{2}_{2}
+(\vartriangle^{2}_{2}-\Omega^{2}_{C}/4)^{2}}}]
\end{eqnarray}
\begin{eqnarray}
Im[\chi]=&&K[f(\theta){\gamma\vartriangle^{2}_{1}\over{\gamma^{2}\vartriangle^{2}_{1}
+(\vartriangle^{2}_{1}-\Omega^{2}_{C}/4)^{2}}}\nonumber\\
&&+g(\theta){\gamma\vartriangle^{2}_{2}\over{\gamma^{2}\vartriangle^{2}_{2}
+(\vartriangle^{2}_{2}-\Omega^{2}_{C}/4)^{2}}}]
\end{eqnarray}
\section{ANALYSIS OF THE SOLUTION}
Eq. (10) and Eq. (11) show that the susceptibility is mainly
determined by $f(\theta)$ and $g(\theta)$.  $f(\theta)$ affects the
dispersion relationship in the vicinity of the resonance with
$\vartriangle_{1}=0$ and $g(\theta)$ affects the relationship near
the resonance at $\vartriangle_{2}=0$. The term of $\cos^{2}\theta$
and $\sin^{2}\theta$ in $f(\theta)$ and $g(\theta)$, denotes the
population of the atoms at the energy level $|1\rangle$ or
$|2\rangle$ respectively and relates with the absorption to the
probe field. While the term of $-\cos\theta\sin\theta$ denotes the
dark state Raman coherence between $|1\rangle$ and $|2\rangle$ and
gives rise Raman amplification to the probe field. The behavior of
these two functions at the range of [0,$\pi/2$] are plotted in FIG.
2.
\begin{figure}
\includegraphics[scale=0.6]{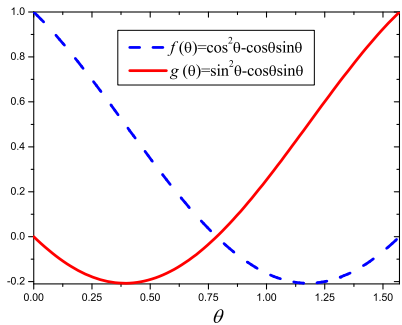}
\caption{\label{fig:wide} Function of $f(\theta)$ (solid line) and
$g(\theta)$ (dash line) in $[0,\pi/2]$ the minimum of the two
expression appear at $\theta=\pi/8$ and $\theta=3\pi/8$}
\end{figure}
\begin{figure} %并排插入两个子图形
\centering \subfigure[]{
\label{fig_2_a} %%标记第一个子图形
\includegraphics[scale=0.4]{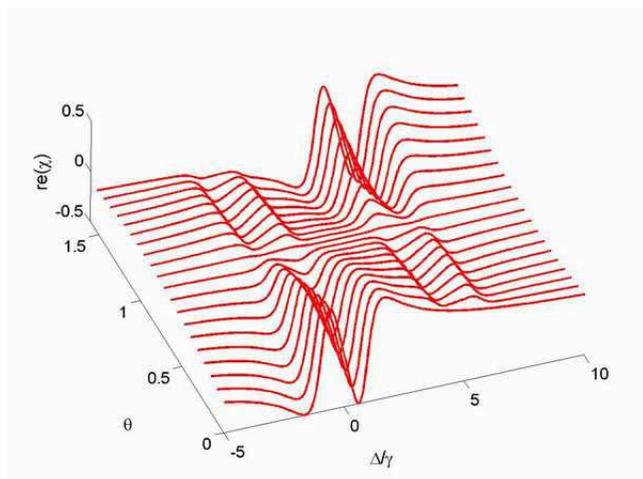}}
\hspace{0.02in} \subfigure[]{
\label{fig_2_b} %%标记第二个子图形
\includegraphics[scale=0.4]{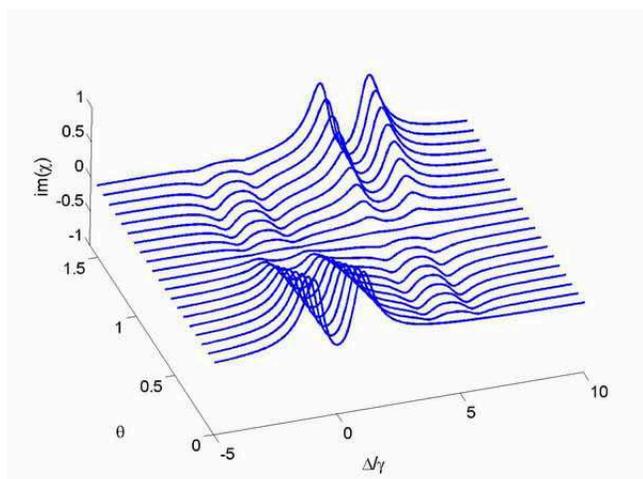}}
\hspace{0.02in} \subfigure[]{
\label{fig_2_c} %%标记第二个子图形
\includegraphics[scale=0.4]{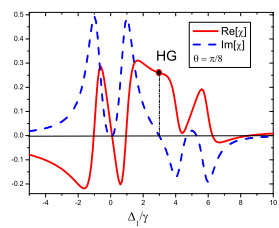}}
\hspace{0.02in} \subfigure[]{
\label{fig_2_d} %%标记第二个子图形
\includegraphics[scale=0.4]{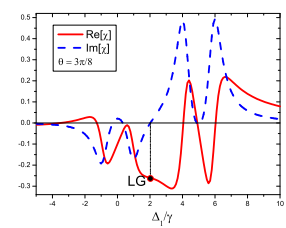}}
\caption{Real and imaginary parts of the susceptibility as a
function of the probe field detuning $\vartriangle_{1}/\gamma$ and
the mixing angle $\theta$. We assume that the material is cold
atomic gas with $N\sim 1.0\times10^{12}$cm$^{-3}$, $\gamma\sim
10$MHz, $\Omega_{C}\approx\ 2\gamma$, $\omega_{21}\approx\ 5\gamma$
and $K\approx\gamma$. (a) and (b) 3D graph about the real part and
the imaginary of the susceptibility. The point of 0 at the axes of
$\vartriangle/\gamma$ is related to the resonance of
$\vartriangle_{1}=0$ and the point of 5 is relate to the resonance
of $\vartriangle_{2}=0$. (c) and (d) Dispersion relationship when
$\theta={\pi/8}$ and $\theta={3\pi/8}$. 'HG' is a high refractive
index without absorption and 'LG' is a negative refractive index
without absorption.} \label{fig_3}
\end{figure}

It is shown from FIG. 2 that $f(\theta)>0$ and $g(\theta)<0$ for
$\theta<\pi/4$. This indicates that the probe experiences absorption
in the vicinity of $\vartriangle_{1}=0$ and gain in the vicinity of
$\vartriangle_{2}=0$. For $\theta>\pi/4$, on the other hand, the
probe experiences gain in the vicinity of $\vartriangle_{1}=0$ and
absorption in the vicinity of $\vartriangle_{2}=0$. Therefor, the
dark state superposition creates mixtures of active and passive
optical materials at frequencies near their resonances\cite{Vikas}.

Particularly, when the mixing angle $\theta=\pi/4$, we have
$\chi(\omega)=0$. In this case, both the real part and the imaginary
part of susceptibility are vanished. Raman coherence of the dark
states induces transparency at all range of frequency, and the probe
field propagates in the medium just as it propagates through the
vacuum.

The minimum of the two functions appear at $\theta=\pi/8$ and
$\theta=3\pi/8$, so that the max gain in vicinity of two resonances
is appearing respectively correspond to $\theta$ at these points. We
plot the real part and the imaginary part of the susceptibility in
FIG. 3. The special points of $\theta=\pi/8$ and $\theta=3\pi/8$ are
also plotted respectively in FIG. 3.

FIG. 3 (a) and (b) show that the population is trapped in level
$|1\rangle$ for $\theta=0$. In the vicinity of $\vartriangle_{1}=0$,
the real and the imaginary part of the susceptibility exhibit a
typical EIT phenomenon in a standard $\Lambda$ system. Similarly,
for the case of $\theta=\pi/2$, the population is trapped in level
$|2\rangle$ and a typical EIT phenomenon is appeared near
$\vartriangle_{2}=0$. When the mixing angle $\theta$ increases from
0 to $\pi/2$, the real part of the susceptibility in vicinity of
resonance at $\vartriangle_{1}=0$ changes from normal dispersion to
abnormal dispersion and contrary similar process can be observed in
vicinity of $\vartriangle_{2}=0$. As the imaginary parts of the
susceptibility near this two resonances are varied proportional to
the function of $f(\theta)$ and $g(\theta)$ respectively.
Especially, one can find that both the real part and the imaginary
part of the susceptibility are viewed as a straight line in these
two figures for the case of $\theta=\pi/4$.

FIG. 3 (c) and (d) show that two points which is labeled as 'HG' and
'LG' . The point of 'HG' has an high refractive index with zero
absorption at $\theta=\pi/8$, while the 'LG' has an negative
refractive index with zero absorption at $\theta=3\pi/8$.
\\
\begin{figure}
\includegraphics[scale=0.65]{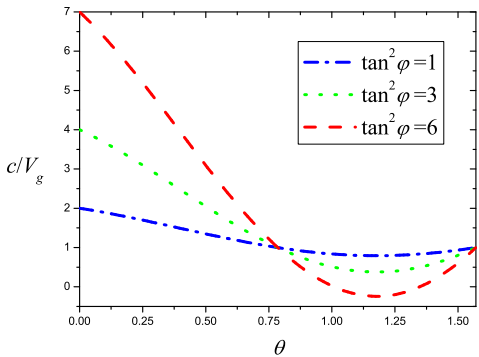}
\caption{\label{fig:wide} Group velocity as a function of $\theta$
($\theta\in[0,\pi/2]$)in some different mixing angle of $\varphi$}
\end{figure}

Assuming a plus with Rabi-frequency $\Omega_{P}(z,t)e^{-i\nu t}$
 propagation in this medium. The Maxwell-Bloch
 equation in one-dimensional slowly-varying envelope approximation is\cite{Scully3}:

\begin{eqnarray}
&&({\partial\over\partial t}+c{\partial\over\partial
z})\Omega_{P}(z,t)=-{\omega_{41}\wp\over
2\epsilon_{0}\hbar}\textmd{Im}(2N\rho_{41}+2N\rho_{42})\nonumber\\
&&
\end{eqnarray}
Let the carrier frequency of the plus $\nu=\omega_{41}$, the effect
of the $N\rho_{42}$ on the right hand side of the Eq. (13) can be
neglected. With the approximation as that adopted in Ref
\cite{Lukin1}, one can obtain that:
\begin{eqnarray}
&&\rho_{41}=-i{2\over\Omega_{C}}{\partial\over\partial
t}\rho_{31}\nonumber\\
&&\rho_{31}\approx-{\Omega_{P}(z,t)\over\Omega_{C}}f(\theta)
\end{eqnarray}
We assume that $\Omega_{C}$ is real and constant, then the
propagation equation is changed to:
\begin{eqnarray}
({\partial\over\partial t}+c{\partial\over\partial
z})\Omega_{P}(z,t)=-{g^{2}N\over\Omega^{2}_{C}}{\partial\over\partial
t}\Omega_{P}(z,t)f(\theta)
\end{eqnarray}
Here $g=\wp\sqrt{2\omega_{41}/\epsilon_{0}\hbar}$, and we define
that $\tan^{2}\varphi=g^{2}N/\Omega^{2}_{C}$. If the mixing angle
$\theta$ is not variable as a function of time. Eq. (15) can be
simplified to:
\begin{eqnarray}
({\partial\over\partial t}+V_{g}{\partial\over\partial
z})\Omega_{P}(z,t)=0
\end{eqnarray}
And
\begin{eqnarray}
V_{g}={c\over{1+f(\theta)\tan^{2}\varphi }}
\end{eqnarray}
It is interesting to compare Eq. (17) to the standard expression of
group velocity in the dispersion media given by:
\begin{eqnarray}
V_{g}={c\over n(\omega)+\omega{dn\over d\omega}}
\end{eqnarray}

Because the carrier frequency which the light field in near
resonance with level $|1\rangle$ and $|4\rangle$, according to the
property of the EIT, the refractive index of the light field
$n(\omega)|_{\vartriangle_{1}=0}=\sqrt{1+Re\{\chi(\omega)\}|_{\vartriangle_{1}=0}}\approx
1$. The comparision shows $\omega({dn\over
d\omega})|_{\vartriangle_{1}=0}\sim f(\theta)\tan^{2}\varphi$. Then
we can draw a conclusion that the sign of $f(\theta)$ determines the
light pulse propagate in the medium for a normal dispersion or an
abnormal dispersion.

For example, at $\theta>\pi/4$, and with near resonant excitation at
$\vartriangle_{1}=0$, we have the normal dispersion. However, when
$\theta>\pi/4$, it is an abnormal dispersion with $f(\theta)<0$ ,
leading to a superluminal light pulse \cite{WLJ}. FIG. 4 shows that
the group velocity as a function of $\theta$ in some of the mixing
angle $\varphi$.

Especially, when
\begin{eqnarray}
\tan^{2}\varphi > |{1 \over {f_{\min}(\theta)}} |\approx 4.83
\end{eqnarray}
the group velocity can become negative. So that one can achieve a
desirable group velocity simply by setting different mixing angle
$\theta$ and $\varphi$ in this system.
\\

An interesting extension of the topic is to vary $\Omega_{1}$ or
$\Omega_{2}$ in the space\cite{Lukin4,ZJY}. This treating leads to a
spatially dependent population distribution at levels $|1\rangle$
and $|2\rangle$ and hence a spatially dependent of the dispersion
relation. If the variation is periodic in the space, it can produce
a photonic band gap (PBG) structure in the
medium\cite{Lukin3,Qiong}.
\begin{figure} %并排插入两个子图形
\centering \subfigure[]{
\label{fig_5_a} %%标记第一个子图形
\includegraphics[scale=0.4]{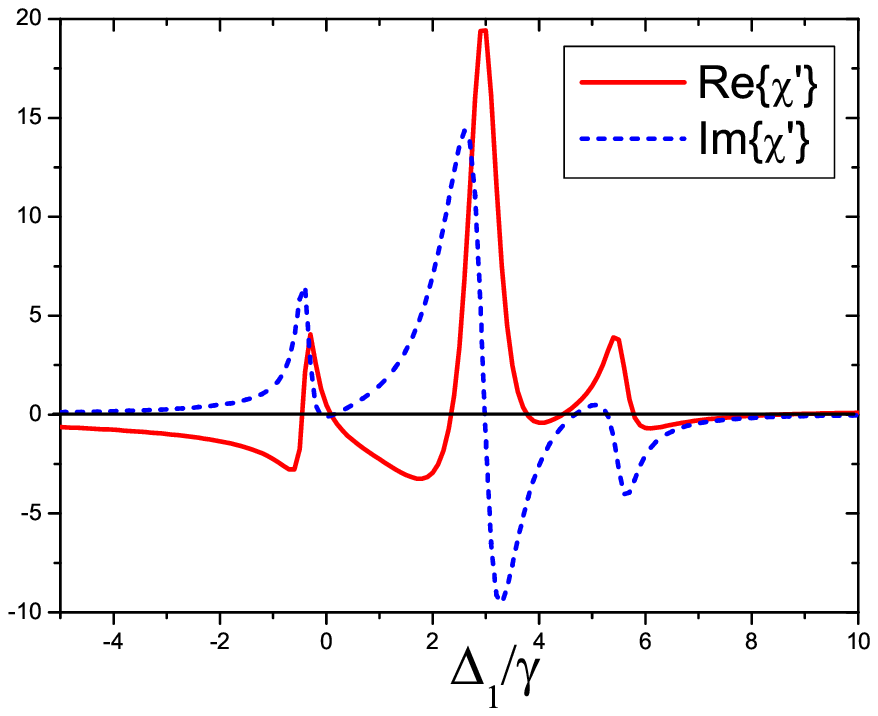}}
\hspace{0.02in} \subfigure[]{
\label{fig_5_b} %%标记第二个子图形
\includegraphics[scale=0.4]{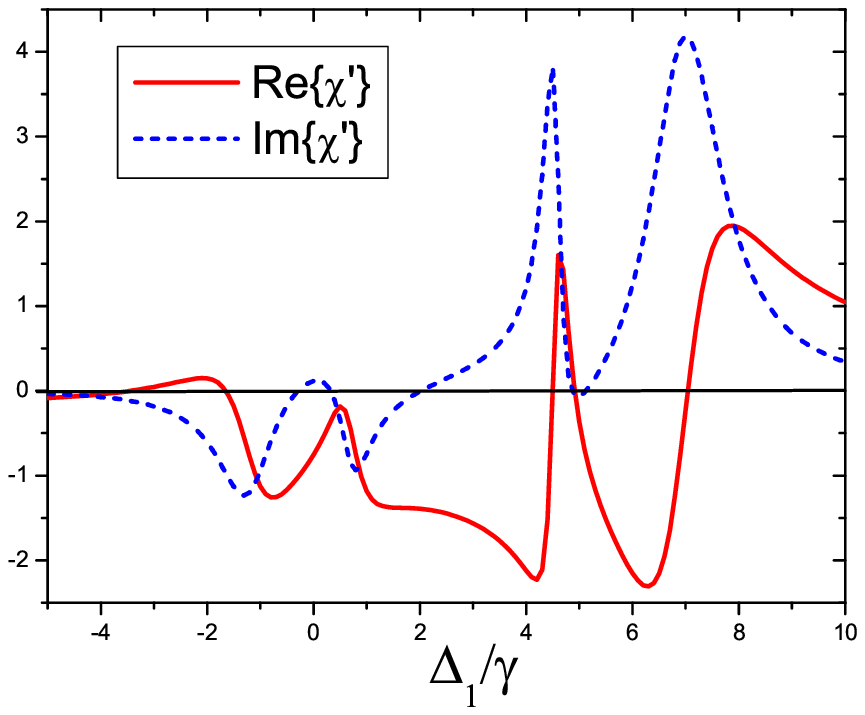}}
\caption{Real and imaginary parts of the susceptibility in the local
field corrections. We chose $N\sim 1.0\times10^{13}$cm$^{-3}$,
$\gamma\sim 10$MHz, $\Omega_{C}\approx\ 2\gamma$,
$\omega_{21}\approx\ 5\gamma$ and $K\approx10\gamma$. (a) and (b):
Dispersion relationship in the local field corrections with
$\theta={\pi/8}$ and $\theta={3\pi/8}$.} \label{fig_5}
\end{figure}

\section{CONCLUSIONS}
Controlled group velocity of light in the tripod system with EIT is
an interesting topic which leads to rich phenomena\cite{Paspalais2}.
In this contribution, coherent superposition of a dark state is
proposed in a four-level tripod system to control the probe
propagation through EIT in the medium. It is found that the dark
state superposition creates mixtures of active and passive optical
materials at frequencies near their resonances and leading to a
controllable  the group velocity either as a sublimunal, a standard,
a superlimunal or even a negative speed. And the propagation
behavior is entirely determined by the mixing angle $\theta$ and
$\varphi$ of the initial states.

The energy levels in the present paper can be established in a wide
range of atomic, molecular, and condensed matter. Rb vapor,
metastable neon, rare-earth atoms, donorbound electrons and bound
excitons in semiconductors\cite{Xiao,GJY,Chen,WYZ,ZY} are
appropriate sample because these medium have abundant level
structures. Ref. \cite{Fleischhauer2007} gives an example\cite{Flik}
of this system in practice. If increasing the density of the
material, the local field effect\cite{Wiser} must take into account.
The electric field replaces to the microscopic field
$E_{m}=E+P/3\epsilon_{0}$ and the dielectric constant is changed as
below:

\begin{eqnarray}
\epsilon-1={N\alpha\over{1-N\alpha/3}}
\end{eqnarray}

where $\alpha$ is the total polarizability. FIG. 5 plots the
susceptibility with $\theta=\pi/8$ and $\theta=3\pi/8$ in the local
field correction.

\begin{acknowledgments}
YongYao Li thanks Dr JingFeng Liu \& GuiHua Chen and Prof WeiMin
Kuang \& Jing Cheng for useful discussion. HuaRong Zhang thanks for
the advice from Prof XiangYang Yu. This work is supported by the
National Key Basic Research Special Foundation (NKBRSF)
(G2004CB719805), Chinese National Natural Science Foundation
(60677051,10774193).
\end{acknowledgments}

\newpage %Just because of unusual number of tables stacked at end
\bibliography{apssamp}% Produces the bibliography via BibTeX.

\end{document}